# The Empirical Commit Frequency Distribution of Open Source Projects

Carsten Kolassa
Software Engineering
RWTH Aachen University,
Germany
Carsten@Kolassa.de

Dirk Riehle
Friedrich-Alexander-University
Erlangen-Nürnberg, Germany
dirk@riehle.org

Michel A. Salim
Friedrich-Alexander-University
Erlangen-Nürnberg, Germany
michel@sylvestre.me

## ABSTRACT

A fundamental unit of work in programming is the code contribution ("commit") that a developer makes to the code base of the project in work. An author's commit frequency describes how often that author commits. Knowing the distribution of all commit frequencies is a fundamental part of understanding software development processes. This paper presents a detailed quantitative analysis of commit frequencies in open-source software development. The analysis is based on a large sample of open source projects, and presents the overall distribution of commit frequencies.

We analyze the data to show the differences between authors and projects by project size; we also includes a comparison of successful and non successful projects and we derive an activity indicator from these analyses. By measuring a fundamental dimension of programming we help improve software development tools and our understanding of software development. We also validate some fundamental assumptions about software development.

## Categories and Subject Descriptors

D.2.8 [**Software Engineering**]: Metrics; D.2.9 [**Software Engineering**]: Management; D.m [**MISCELLANEOUS**]

## 1. INTRODUCTION

Free/libre/open source software (FLOSS) has been adopted widely by industry in recent years. In 2008, 85% of all enterprises used open source software [8]. A 2010 study estimates that 98% of all enterprises use open source software [22].

Given the significance of open source and the interest in the field of mining software repositories [13], it is surprising that there are few statistical analyses that cover a *representative* percentage of open source.

As open source software is being used for critical tasks, it is important to have a quantitative analysis of as broad a range of open source projects as possible, so that research can characterize their development methodologies in an informed way.

Previous analyses are limited in two ways: they tend to be case studies focusing either on key open source projects, or on a small sampling of projects; and they only briefly mention commit frequencies (e.g. [7], [2], [4]) or don't mention commit frequency at all.

This work addresses these two limitations: our database covers approximately 30% of all open source software projects of its time and we focus exclusively on commit frequencies. Our companion paper, [15], likewise focuses on commit sizes. The size of our dataset also allows us to empirically show certain general patterns like the "daily commit" or the commit shortly after one that broke the build can be statistically proven. We also show that ther is a correlation between activity and successfulness of a project.

This combination of narrow focus and breadth of data allows us to analyze properties that have not been analyzed before: for instance, tracking the activities of individual developers across projects, as well as deriving more broadly applicable conclusions from the properties observed in previous studies (e.g. the prevalence of commits by top developers in individual projects). This narrow but deep focus furthermore allows us to do a cross project comparison to investigate if the commit frequency can be used as an indicator for project performance.

The contributions of this paper are:

- the empirical commit frequency distribution of open source software projects,
- an analysis of the differences between developers,
- the empirical distribution of commits of a developer to a particular project, and
- an analysis of the differences between the commit frequency distributions of projects of different sizes, as well as
- a comparison of successful and unsuccessful projects and an
- activity indicator for projects.

The paper is organized as follows. Section 2 defines some terms we use. Section 3 contains our analysis of commit frequencies. Section 4 discusses threats to validity. Section 5 discusses related work, and section 6 concludes the paper.

## 2. COMMIT FREQUENCIES

A software project is typically developed in multiple iterations, in a series of changes to its artifacts, for instance code,

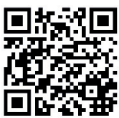

documentation and artwork. When the project is managed using a *version control system* (VCS; also known as a *source configuration management* system), these changes are organized into sets known as *commits*.

From this range of potential artifacts, in the following, we focus on commits that involve source code contributions, and refer to them as commits.

We define *commit frequency* as the number of commits in a time period; this, however, is not directly measurable; thus we first measure the *commit interval* which is the time between two commits, and derive a commit frequency measure from it.

## 3. MEASUREMENTS

### 3.1 Data Source and Research Method

This paper uses the database of the Ohloh.net open source project index. Our database snapshot is dated March 2008. It contains 11,143 open source projects with a total of 8,705,118 commits by 47,548 committers. Daffara estimates that there were 18,000 active open source projects in September 2007 worldwide [5]. The total number of projects is much larger, but most open source projects are not active and by our activity definition have to be excluded. We use the same definition of "active project" as Daffara: A project is active at a given point of time when the number of commits in the preceding 12 months is at least 60% of the number of commits in the 12 months before that. Using this definition our data set contains 5,117 active open source projects. We therefore estimate that our database contains about 30% of all open source projects considered active in March 2008.

Our analysis is descriptive: we are discovering existing characteristics in the data rather than starting off with a hypothesis and attempting to invalidate or validate it. We also split our analysis along project sizes and provide the characteristics of commit frequency by project size. Using the Ohloh data, we calculated the interval between two commits. We do this, because the commit interval is not prone to timezone errors. Thus it is always calculated for a single committer. To be able to track this committer we used Ohloh.net's committer identities which allow us to have a committer ID across project boundaries.

### 3.2 Commit Frequency Distribution

The commit frequency is the number of commits in a given time span. Thus it is the inverse of the commit interval which we defined as the time between two commits.

The number of occurences of a certain commit interval results in the commit interval distribution. We measure it instead of the equivalent but not directly measurable commit frequency distribution.

The distribution of those commit intervals shows how likely it is that a certain time elapses between two commits of the same committer.

In statistics a distribution can be represented as a *probability distribution function* (PDF) or a *cumulative distribution function* (CDF). The PDF in our case describes the relative likelihood that a particular time interval lapses between two commits of the same committer.

The CDF can be computed by integrating the PDF. Integrating the PDF over an interval provides the probability that the time that lapses between two commits is within a certain interval. For example, integrating over the interval $[0s, 10min]$ provides the probability that the time between two commits of the same committer is between 1 second and 10 minutes.

| Key Parameter | Value |
|:---:|:---:|
| Median | 1.666 h |
| Mean | 3.206 d |
| 90th percentile | 4.075 d |
| 95th percentile | 9.427 d |

**Table 1: Statistical key characteristics of the open source commit intervals in open source.**

Neither the PDF nor the CDF are sensitive to the absolute time (timezone etc.) of the committer as we only measure the time between commits; thus this time offset has no impact.

The empirical result of our measurements is the *empirical probability distribution function* (EPDF) as shown in figure 1. The EPDF is a density estimation based on the observed data. The EPDF is not a closed model, it is a representation of the observed data. The statistical key characteristics of commit intervals in open source are shown in table 1.

To make it easier to visually inspect the PDF, we have added a plot showing only the density estimates for commits having an interval of less than 24 hours (see figure 2).

Integrating the EPDF yields the *empirical cumulative distribution function* (ECDF). Using the ECDF we can determine that the time between two commits of the same committer is normally less than one day with 50% of the commits happening in less than 1.666 h after a former commit and that there are local maxima every 24 hours. Those 24 hours maxima correspond to commits which have approximately 24 hours between them.

We not only present the EPDF and ECDF but also table 2 which shows the percentage of all commits happening in a certain time interval after a preceding commit.

### 3.3 Comparison by committer

In the previous section, after computing the commit intervals for individual committers, we provided aggregate measures for the combined commit intervals of all committers. We now compare committers to each other to determine differences in commit frequency from committer to committer. We do this by calculating the median commit interval for each committer. The results show that the vast majority of committers works very regularly on open source projects with 50% of the committers having a median commit interval of less than 13.78 h (for other statistical key characteristics see table 4). We present the EPDF of the distribution of the median commit interval of committers in figure 3. It shows a tendency towards short commit intervals clearly but it also shows local maxima on multiples of 24 hours.

We also look at the mean commit interval which is much larger because it incorporates outliers (huge commit intervals, e.g. because a committer was on vacation). Based on this we calculate the mean number of commits per week and the mean number of commits per day. We use the mean instead of the median to calculate this aggregated commit frequencies because we cannot say for sure if those outliers are always created because of external events like vacation or if a committer just needed more time for this particular commit. The mean number of commits per day and per week for the different percentiles of developers are shown in table 3.

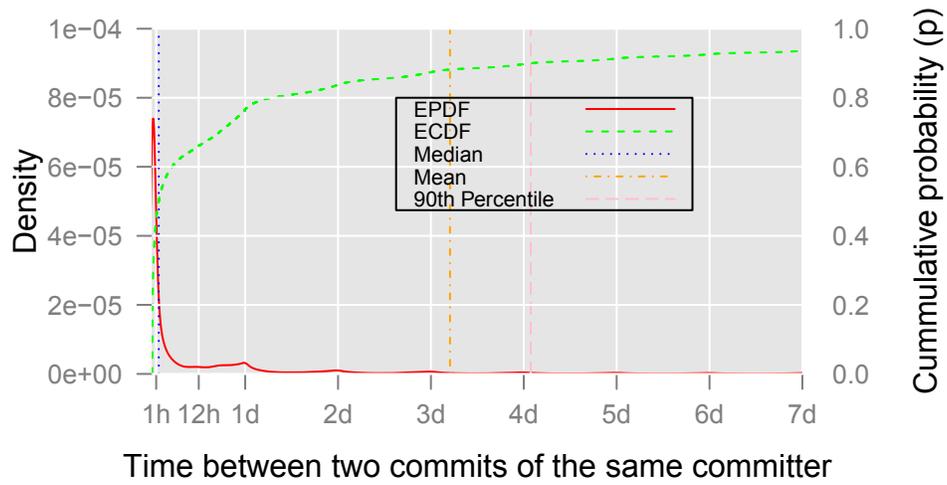

Figure 1: The EPDF and ECDF of time between commits (up to an interval of 1 week)

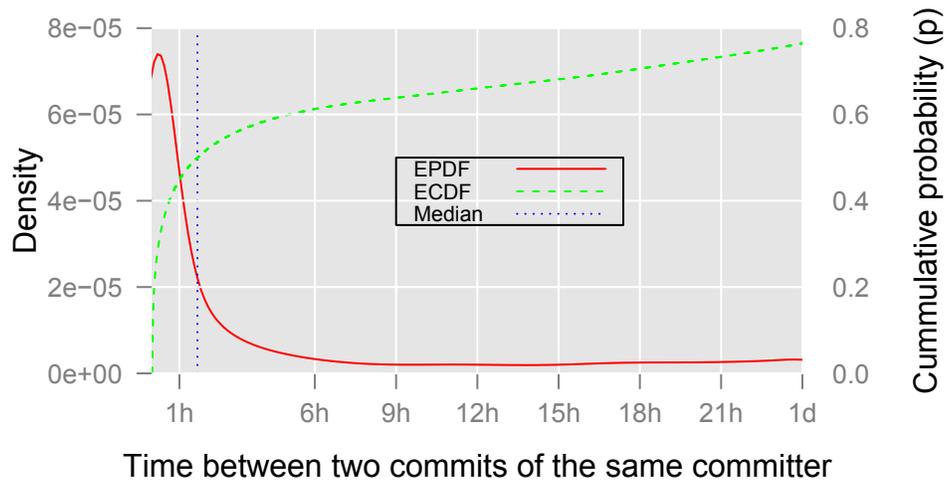

Figure 2: The EPDF and ECDF of time between two consecutive commits of (up to an interval of 1 day)

| Time between two commits of the same committer | 0h-1h | 1h-2h | 2h-3h | 3h-4h | 4h-5h |
|---|---|---|---|---|---|
| Occurences in percent | 44.8 | 7.0 | 3.8 | 2.5 | 1.8 |
| Time between two commits of the same committer | 5h-6h | 6h-7h | 7h-8h | 8h-9h | 9h-10h |
| Occurences in percent | 1.3 | 1.0 | 0.8 | 0.7 | 0.7 |

Table 2: Mean time between commits in percent.

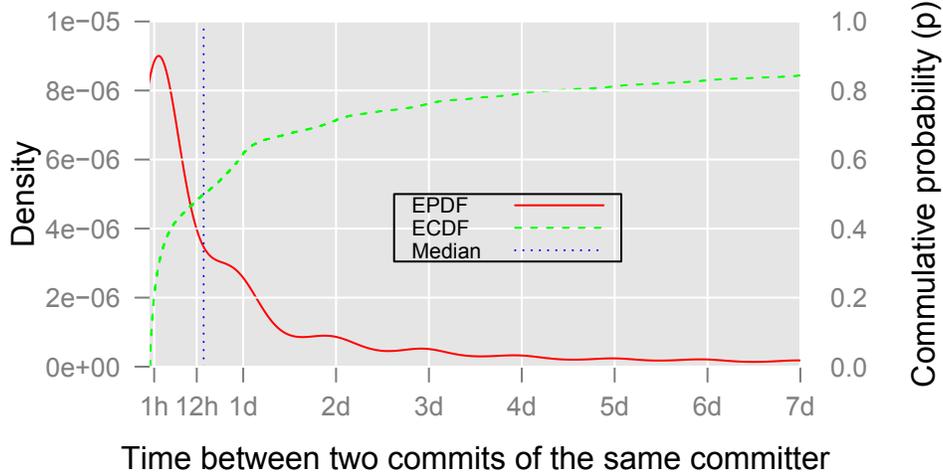

Figure 3: The EPDF and ECDF of median commit intervals per committer.

|  | 0%-10% | 10%-20% | 20%-30% | 30%-40% | 40%-50% |
|---|---|---|---|---|---|
| Average commits per week | 28.07 | 6.67 | 3.66 | 2.32 | 1.51 |
| Average commits per day | 4.01 | 0.95 | 0.52 | 0.33 | 0.22 |
|  | 50%-60% | 60%-70% | 70%-80% | 80%-90% | 90%-100% |
| Average commits per week | 0.99 | 0.63 | 0.38 | 0.20 | 0.04 |
| Average commits per day | 0.14 | 0.09 | 0.05 | 0.03 | 0.01 |

Table 3: Mean commit frequency per committer by the percentile of commit frequency

| Key Parameter | Value |
|---|---|
| Mean | 13.62 d |
| Median | 13.78 h |
| 90th percentile | 16.51 d |
| 95th percentile | 51.4 d |

Table 4: Statistical key characteristics of the distribution of the median of the commit intervals for particular authors.

### 3.4 Comparison by committer and project

We not only examine the commit frequency of committers across all projects the particular committer is working on. We also look at committer/project pairs and examine the commit frequency of a particular committer to a particular project.

We classify the projects into small, medium, and large size projects based on the number of involved developers. Becher provides an analysis of the number of developers in a random sample of projects included in the Debian GNU/Linux distribution [3]. We use their proposed partitioning to group our projects accordingly (see table 5).

Figure 4 and figure 5 shows that the divergences between the project categories are very small. In fact, a close examination of the ECDFs shows that the maximum difference is only 6.3%. We posit that these dissimilarities are caused by differences in commit management. A large project typically has contributors that commit patches from different committers, but in the Ohloh database those differences between committer and author are not represented; this leads to a very small commit interval for these committers.

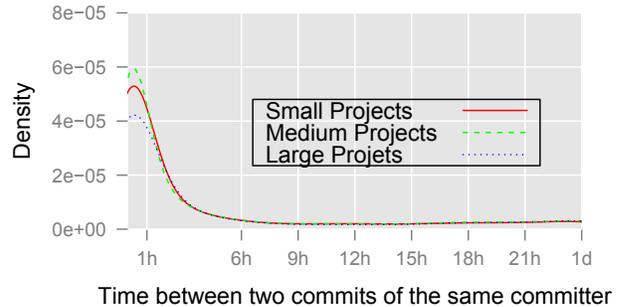

Figure 4: Combined EPDF plot by project size

| Parameter | Minimum number of developers | Maximum number of developers |
|---|---|---|
| Small | 1 | 5 |
| Medium | 6 | 47 |
| Large | 48 | $\infty$ |

Table 5: Project size boundaries

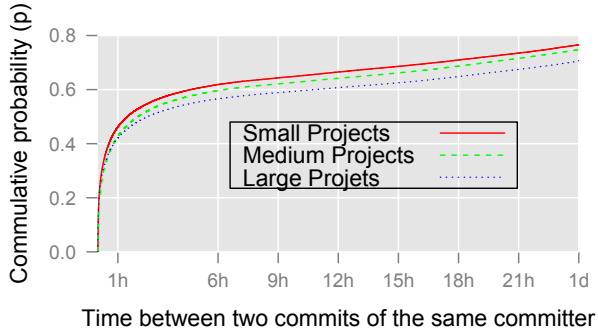

Figure 5: Combined ECDF plot by project size

### 3.5 Comparison by project success

Another question is how the commit frequency of committers in successful projects differs from those in unsuccessful ones. To answer this question we first have to come up with a definition of success. There are many ways to define success in open source projects. The two most common are:

**Popularity**: Popularity describes whether the project is used and liked by the open source community. We use the ohloh popularity index to decide whether a project is popular or not. The ohloh popularity index is equal to the number of web pages linking to the project home page as determined by the Yahoo search engine API [1].
**Survivability**: It describes whether a project stays active. To analyze this we divided the projects in two groups active ones and inactive ones and compare the commit frequency of active projects to the commit frequency of inactive ones to do this we use the same definition of "active project" as Daffara as mentioned in section 3.1.

Popularity and project activity are mentioned as success indicator for open source projects in [19] while [6] only mentions popularity.

To compare popular projects to non popular ones we compare the 100 most popular projects in our database to the rest of the projects. We see that unpopular projects have a much higher chance of high commit frequencies for a single commiter. Figure 6 shows the distribution of time between commits of the same author to the same project and figure 7 shows the median commit frequencies of the projects. We posit that the higher commit frequency in small projects is caused by the much tougher review process in popular projects which causes the delay [20].

To show that the commit frequency is an indicator for project activity. We calculate the median commit frequency in a project over the last 6 months and over the whole lifetime of the project and then we calculate the ratio of those two.

When a project becomes inactive the developers commit less and therefore the time between commits increases, and thus the commit frequency is a good indicator for the project activity. Figure 8 shows the distribution of the ratios comparing active and inactive projects classified using the Daffara criteria.

At a threshold of 0.47 the difference to the Daffara criteria is the smallest. 24.4% of the projects that Daffara classifies as active have been classified as inactive by our metric. This difference sounds bigger than it is. We compare two metrics to each other if they were exactly the same our metric would be redundant. The less active a project is the longer the time between commits of individual developers.

Therefore it is logical to use the commit frequency as an activity metric. The clear maxima in figure 8 show that for projects approaching inactivity the metric gets closer and closer to zero while for projects that are active the commit frequency stays roughly the same or even increases. Thus it makes sense to use it as an indicator in a survivability metric, which indicates if a project is still healthy and active or whether steps need to be taken because the project became less active.

## 4. THREATS TO VALIDITY

### 4.1 Committer vs author

While ideally the two definitions coincide exactly, in practice this is not always the case. In open source there are committers which commit patches that are developed by other developers. But there are also committers that only commit patches they developed themselves. We cannot reliably distinguish the one from the other. We just track the time between commits of individual committers. We believe that this is not problematic as it simply reflects the actual commit behavior of committers.

### 4.2 Invalid timestamps

There are outlier timestamps, either newer than the database itself (as far in the future as the 22nd century) or older than most version-control systems.

The number of outliers is very small – less then one-thousandth, so we believe that they don't have any significant impact. Thus, we removed them.

### 4.3 Different IDs in different Projects

The commiters could have different IDs in different Projects. We use the author id from the ohloh.net database which is derived from the email address. If committers use different email addresses in different projects this can lead to different IDs and therfore distort our results.

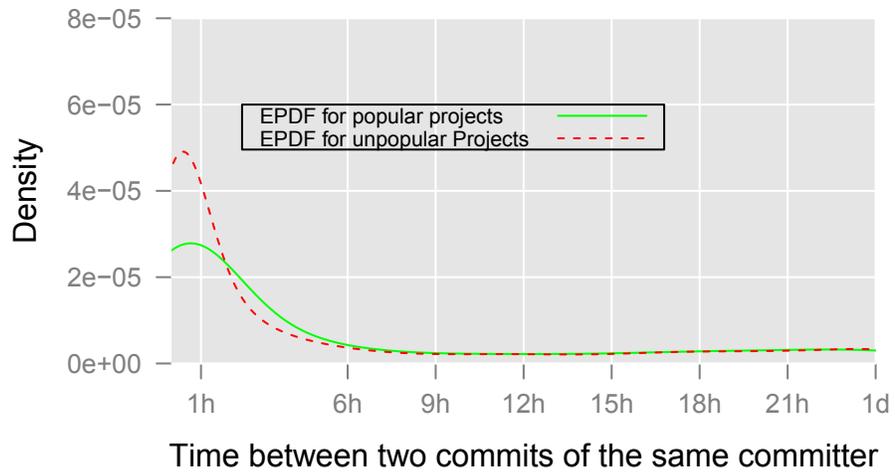

Figure 6: The EPDF of the time between commits of popular and unpopular projects

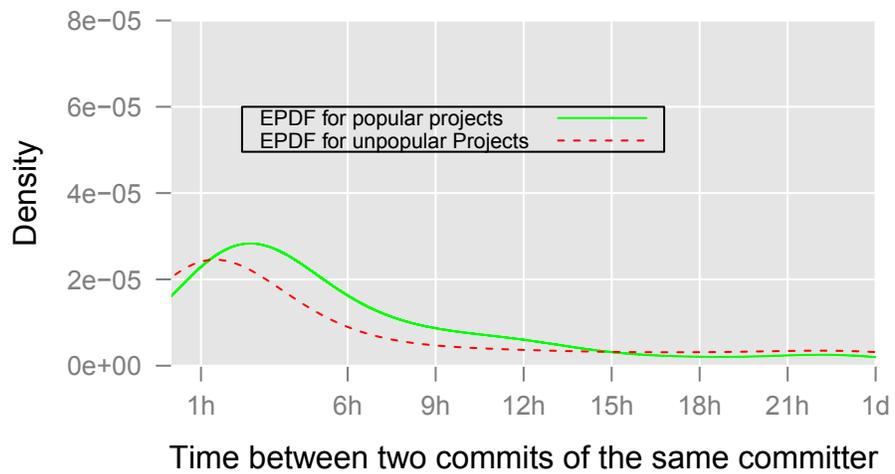

Figure 7: The EPDF of the median time between commits of popular and unpopular projects

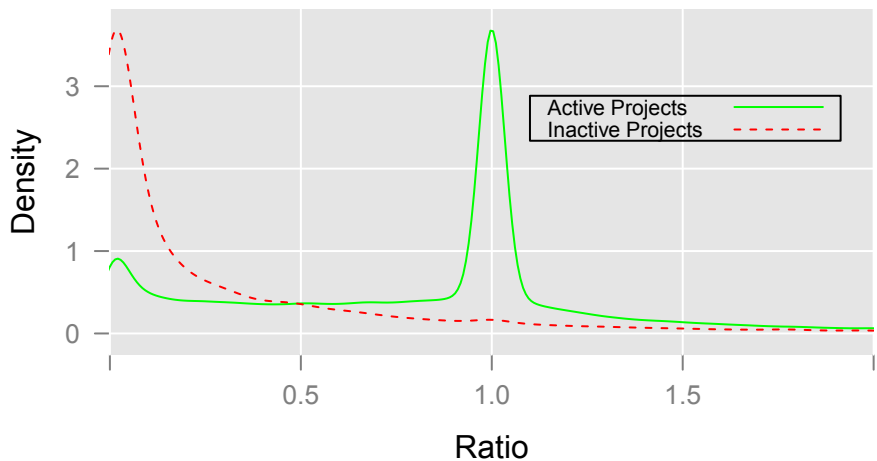

Figure 8: The EPDF of the ratio between the median commit frequency in the last 6 month of an analyzed month and the project lifetime before that month

### 4.4 Project rules for commit strategies

Some projects have rules for commit strategies e.g. no commits of new features close to a release date the so called feature freeze. We see these commit strategies as part of the general commit frequency distribution. It is true that these strategies have an impact and we reproduce this impact.

## 5. RELATED WORK

After a thorough review of related work, we found no previous research on this aspect of open source programming behavior. Commit frequencies are briefly mentioned in Daniel M. German's work on mining CVS repositories.

German reconstructs "modification requests", i.e. commits, from source code repositories of selected projects, where modifications are recorded on a per-file basis [11, 10, 9]. German provides a *modification request per month* in passing [9], and the term *modification frequency* is defined in [12]. [14] analyses commits and tries to find reoccuring events using fourier analysis. But the papers don't provide a commit frequency distribution or detailed statistical data about commit frequencies.

[16] briefly mentions the time it takes to fill modification requests, this is the time it takes till a change to the code is commited from the time of the request to the actual commit. In contrast to the commit interval which is the time between two commits.

[17] uses changes per hour as a measure for code changes but the changes are not commits in this case but rather every change in source code (e.g. renaming a function is one change) this leads to very high numbers of changes per hour (between 25 and 46 changes per hour) the paper has therefore a completely different scope.

[23] studies how many commits conflict but the time between commits or how often developers commit was not investigated.

[18] and [21] focus on predicting where in the code a change occurs but don't address when a change occurs.

## 6. CONCLUSIONS

We found that committers have small commit frequencies with an average time between commits of 3.206d. But we also found out that this is most likely due to larger pauses in development because the median commit frequency of the median author is higher, in fact 50% of all authors have less than 13.78 h between 50% of their commits. A possible explanation is that many authors try to slice their contributions into several commits or because committers need to hotfix a commit because it breaks someone's build, or has other side-effects, or bugs. Both are reasons for having two commits in a short period of time. The difference between the median of the commit interval over all commits and the median of individual commit intervals shows that there is a group of committers that committs an order of magnitude more often that the average committer this group most likely has the comitter role in projects. We also observed maxima in the EPDF every 24 hours which are caused by committers that commit a "daily commit" roughly on the same time every day. Another important discovery is that we've presented in this paper is that small, medium and large project only differ slightly in terms of commit frequency with a maximum difference between the ECDFs of 6.34%.

Finally, knowing the commit frequency distribution allows us to better design and provision configuration management systems. Using our results, we can better predict the estimated time of arrival of the next commit to a configuration management system and implement it accordingly. Also, knowing the number of users and using our results lets us predict the workload when provisioning and operating a configuration management system. Thus, our results should be helpful to configuration management users, developers, and operators alike.

The empirical knowledge gained from actually measuring the commit frequency distribution is an important first step to create hypotheses for research. But more important it can be used as a benchmark to compare projects and developers. A commit is the smallest increment a developer contributes to the code base of a project therefore the commit frequency distribution allows us to rank the developer by how frequently he makes a contribution. This in turn allows us to compare a developer with an average developer and it furthermore allows us to give an exact percentile ranking for an individual developer. With this information we can determine whether he has a high commit frequency that makes agile development easier as it prevents merge conflicts on the repository or not.

But we can not only compare developers to each other, we can also compare projects. As shown we can compare them to the overall commit frequency distribution to get a percentile ranking. But we can also compare a projects commit frequency statistic to itself to determine weather it is active or not. In short the commit frequency is a fast indicator to determine if the project is healthy because it has regular contributions and if the developers are productive by checking whether they contribute regularly.